\input phyzzx
\def\Im{\mathop{\rm Im}\nolimits}
\REF\GUI{%
{\sl Large-Order Behavior of Perturbation Theory}, J. C. Le~Guillou
and J. Zinn-Justin eds., (North-Holland, Amsterdam, 1990).}
\REF\LIP{%
L. Lipatov,
JETP Lett. 25 (1977) 104; Sov.\ Phys.\ JETP 45 (1977) 216.}
\REF\BRE{%
E. Br\'ezin, J. C. Le Guillou and J. Zinn-Justin,
Phys.\ Rev.\ D15 (1977) 1544; D15 (1977) 1558.}
\REF\LAU{%
B. Lautrup,
Phys.\ Lett.\ 69B (1977) 109.}
\REF\THO{%
G. 't Hooft,
in {\sl The Whys of Subnuclear Physics}, A. Zichichi ed., (Plenum,
New York, 1978).}
\REF\OLE{%
P. Olesen,
Phys.\ Lett.\ 73B (1978) 327.}
\REF\PAR{%
G. Parisi,
Phys.\ Lett.\ 76B (1978) 65; Phys.\ Rep.\ 49 (1979) 215.}
\REF\BER{%
M. C. Berg\`ere and F. David,
Phys.\ Lett.\ 135B (1984) 412.}
\REF\BEN{%
M. Beneke,
Nucl.\ Phys.\ B405 (1993) 424;
M. Beneke and V. A. Smirnov,
Nucl.\ Phys.\ B472 (1996) 529.}
\REF\VAI{%
A. I. Vainshtein and V. I. Zakharov,
Phys.\ Rev.\ Lett.\ 73 (1994) 1207; 75 (1995) 3588; Phys.\ Rev.\ D54
(1996) 4039.}
\REF\LOV{%
C. N. Lovett-Turner and C. J. Maxwell,
Nucl.\ Phys.\ B432 (1994) 147; B452 (1995) 188.}
\REF\CEC{%
G. Di Cecio and G. Paffuti,
Int.\ J. Mod.\ Phys.\ A10 (1995) 1449.}
\REF\REV{%
For review:
R. Akhoury and V. I. Zakharov,
Nucl.\ Phys.\ Proc.\ Suppl.\ 54A (1997) 217.
M. Beneke, hep-ph/9706457, and references therein.}
\REF\GRO{%
D. J. Gross and A. Neveu,
Phys.\ Rev.\ D10 (1974) 438.}
\REF\PARI{%
G. Parisi,
Nucl.\ Phys.\ B150 (1979) 163.}
\REF\DAV{%
F. David,
Nucl.\ Phys.\ B209 (1982) 433; B234 (1984) 237; B263 (1986) 637.}
\REF\MUE{%
A. H. Mueller,
Nucl.\ Phys.\ B250 (1985) 327.}
\REF\LEE{%
T. Lee,
Phys.\ Rev.\ D56 (1997) 1091.}
\REF\SEI{%
N. Seiberg,
Phys.\ Lett.\ B206 (1988) 75, and references therein.}
\REF\SEIB{%
N. Seiberg and E. Witten,
Nucl.\ Phys.\ B426 (1994) 19; B431 (1994) 484.}
\REF\REVI{%
For review:
A. Bilal,
hep-th/9601007.
R. Flume, L. O'Raifeartaigh and I. Sachs,
hep-th/9611118.
L. Alvarez-Gaum\'e and S. F. Hassan,
Fortsch.\ Phys.\ 45 (1997) 159, hep-th/9701069.}
\REF\FIN{%
D. Finnell and P. Pouliot,
Nucl.\ Phys.\ B453 (1995) 225.}
\REF\ITO{%
K. Ito and N. Sasakura,
Phys.\ Lett.\ B382 (1996) 95; Nucl.\ Phys.\ B484 (1997) 141; Mod.\
Phys.\ Lett.\ A12 (1997) 205.}
\REF\DOR{%
N. Dorey, V. V. Khoze and M. P. Mattis,
Phys.\ Rev.\ D54 (1996) 2921; D54 (1996) 7832; Phys.\ Lett.\ B388
(1996) 324; B390 (1997) 205; Nucl.\ Phys.\ B492 (1997) 607.}
\REF\AOY{%
H. Aoyama, T. Harano, M. Sato and S. Wada,
Phys.\ Lett.\ B388 (1996) 331.
T. Harano and M. Sato,
Nucl.\ Phys.\ B484 (1997) 167.}
%
\FIG\figone{%
Sum of strings of one loop bubbles,~$i\Delta(p^2)$.}
\FIG\figtwo{%
$O(1/N)$ correction to the self energy
part,~$\Sigma_{ij}^{(2)}(p^2)$.}
\FIG\figthree{%
$O(1/N^2)$ corrections to the four point vertex
function,~$\Gamma_{ijkl}^{(2)}$.}
\pubnum={%
IU-MSTP/26; hep-th/9710013}
\date={October 1997}
\titlepage
\title{%
Renormalon's Contribution to Effective Couplings}
\author{%
Hiroshi Suzuki\foot{%
electric mail: hsuzuki@mito.ipc.ibaraki.ac.jp}}
\address{%
Department of Physics, Ibaraki University, Mito 310, Japan}
\abstract
When an asymptotically non-free theory possesses a mass parameter,
the ultraviolet (UV) renormalon gives rise to non-perturbative
contributions to dimension-four operators and dimensionless
couplings, thus has a similar effect as the instanton. We illustrate
this phenomenon in $O(N)$~symmetric massive $\lambda\phi^4$~model in
the $1/N$~expansion. This effect of UV renormalon is briefly compared
with non-perturbative corrections in the magnetic picture of the
Seiberg-Witten theory.
\endpage
At present, there are two known sources which make the perturbation
series in quantum field theory divergent~[\GUI]. One is the
instanton~[\LIP,\BRE], Euclidean classical solution with a finite
action, and another is the ultraviolet (UV)~[\LAU--\REV] and the
infrared~[\THO,\BEN,\LOV--\LEE] renormalons. In this letter, we
point out that renormalons and instantons are also similar in another
aspect: In an asymptotically non-free theory, UV renormalon gives
rise to ``non-perturbative'' contributions to dimension-four
operators, when the theory possesses a mass parameter or scale. In
the context of supersymmetric gauge theories, a similar effect of
instanton was pointed out in~[\SEI] and the instanton contributions
in $N=2$~supersymmetric gauge theory was determined to all orders in
the recent work of Seiberg and Witten~[\SEIB,\REVI].

The renormalon is a Feynman diagram which has the amplitude grows
like~$\sim n!$ for the $n$th~order loop expansion. In asymptotically
non-free theories, a singular UV behavior of Feynman integral due to
the Landau pole produces this factorial growth with a non-alternating
sign. After the Borel resummation, this large order behavior produces
a contribution proportional to~$\sim e^{-n/(\beta_1\alpha)}$
($\alpha$~is a typical coupling constant and $\beta_1$~is the one
loop coefficient of the $\beta$~function) which
is~$O(1/\Lambda^{2n})$ in terms of the $\Lambda$~parameter.
Therefore, if there exists a mass parameter, UV renormalon can give
rise to ``non-perturbative'' corrections of the
order~$O(m^{2n}/\Lambda^{2n})$ to four-dimensional operators.

We illustrate this phenomenon in $O(N)$~symmetric massive
$\lambda\phi^4$~model,
$$
   {\cal L}={1\over2}\partial_\mu\phi_i\partial^\mu\phi_i
   -{1\over2}m^2\phi_i^2-{\lambda\over8N}({:}\phi_i^2{:})^2,
\eqn\one
$$
where $\phi_i$ is an $N$~component scalar field ($i=1$--$N$).
In~\one, we have introduced the normal ordering to suppress (most of)
tadpole diagrams and to avoid unnecessary complications associated
with the dynamical mass generation. Also, the normalization of the
coupling constant has been taken in accord with the standard
$1/N$~expansion, because the $1/N$~expansion allows a systematical
isolation of renormalon diagrams, as is well-known. In what follows,
we shall study the self energy part and the four point vertex
function to the next-to-leading order of the $1/N$~expansion.

In the lowest order of the $1/N$~expansion, it is easy to see that
there is no self-energy correction (with the normal ordering),
$\Sigma_{ij}^{(1)}(p^2)=0$, and thus the mass parameter~$m$ in~\one\
is the physical one. On the other hand, in the lowest order of the
$1/N$~expansion, the four point vertex function at the zero external
momentum~$\Gamma_{ijkl}(0)$ is given by
$$
   \Gamma_{ijkl}^{(1)}(0)=-{1\over N}
   (\delta_{ij}\delta_{kl}+\delta_{ik}\delta_{jl}
    +\delta_{il}\delta_{jk})i\Delta(0),
\eqn\two
$$
where $i\Delta(p^2)$ is a geometrical sum of strings of one-loop
bubbles in~Fig.~\figone,
$$
   i\Delta(p^2)=\sum_{n=0}^\infty\Pi(p^2)^n\lambda^{n+1}
   ={1\over{1/\lambda}-\Pi(p^2)},
\eqn\three
$$
and $\Pi(p^2)$~is the {\it renormalized\/} one loop bubble defined
by\foot{%
Throughout this letter, the symbol~$e$ is used only for the Napier's
number, $e=2.718\cdots$.}
$$
\eqalign{
   \Pi(p^2)
   &=-{\beta_1\over2}\int_0^1dx\,\ln{e^C\mu^2\over m^2-p^2x(1-x)}
\cr
   &=-{\beta_1\over2}\left[\ln{e^C\mu^2\over-p^2}+2
       +2{m^2\over-p^2}\left(\ln{m^2\over-p^2}-1\right)
       +O\left(\left({m^2\over-p^2}\right)^2\ln{m^2\over-p^2}\right)
      \right],
\cr
}
\eqn\four
$$
($\mu$~is the renormalization point). The expression of the
renormalized bubble~\four\ depends on the renormalization scheme, but
the dependence can be parameterized by a single constant~$C$~[\BEN].
For example, we have $C=0$~in the $\overline{\rm MS}$~scheme and,
$C=-\gamma+\ln4\pi$ in the MS scheme. The coefficient of the
renormalized bubble is related to the renormalization group (RG)
$\beta$~function of~$\lambda$ as
$$
   \beta(\lambda)=\mu{\partial\over\partial\mu}\lambda
                 =\beta_1\lambda^2+O(1/N),
\eqn\five
$$
and~$\beta_1=1/16\pi^2>0$.

We define the RG invariant $\Lambda$~parameter as the value of the
renormalization point~$\mu$ at which the running
coupling~$\lambda(\mu)$ diverges. Therefore, in any renormalization
scheme, we have~$\Lambda=e^{1/(\beta_1\lambda)}\mu$ to this order of
the $1/N$~expansion. On the other hand, $i\Delta(p^2)$~in~\three\ is
an S-matrix element and thus must be independent of the
renormalization scheme. This shows a combination~$e^{C/2}\Lambda$ is
scheme-independent. For simplicity, we will omit the symbol~$C$ in
the following because the dependence on~$C$ can be recovered by
substitutions, $\mu^2\to e^C\mu^2$ and~$\Lambda^2\to e^C\Lambda^2$.
In particular, our final results (13)--(17) are independent of the
renormalization scheme after this substitution.

In usual analyses of UV renormalon, the~$O(m^2/p^2)$ or higher order
terms in~\four\ are neglected, because the contribution of UV
renormalon arises from the Euclidean UV singularity of~\three\
at~$p^2=-\Lambda^2$, i.e., at the Landau pole, and $m\ll\Lambda$ for
a weak coupling~$\lambda\ll1$. When these higher order terms
in~\four\ are neglected (thus the theory is basically regarded as a
massless one), the leading UV renormalon contribution emerges in
dimension six operators~[\PAR,\BER]. However, the $O(m^2/p^2)$~term
in~\four\ is crucial and cannot be neglected for our purpose to find
$O(m^2/\Lambda^2)$~corrections to dimension four operators.

Let us next consider the next-to-leading order corrections in the
$1/N$~expansion, the first non-trivial order the UV renormalon
emerges. In this order, the self-energy correction and the four-point
vertex function are respectively given by the diagrams in
Fig.~\figtwo\ and~Fig.~\figthree. In these figures, the double line
denotes the sum of strings of bubbles in~Fig.~\figone. At this stage,
it is convenient to introduce the Borel representation of the sum of
strings of bubbles~[\PAR]:
$$
   i\Delta(p^2)=\int_0^\infty dz\,e^{-z/\lambda}B(z;p^2).
\eqn\seven
$$
Then, from \three\ and~\four, we see that the Borel
transform~$B(z;p^2)$ has the following structure:
$$
\eqalign{
   &B(z;p^2)
   =\sum_{n=0}^\infty{1\over n!}\Pi(p^2)^nz^n
   =\exp\left[\Pi(p^2)z\right]
\cr
   &=e^{-\beta_1z}\left({\mu^2\over-p^2}\right)^{-\beta_1z/2}
   \left[1-{m^2\over-p^2}\left(\ln{m^2\over-p^2}-1\right)
   \beta_1z
   +O\left(\left({m^2\over-p^2}\right)^2\ln{m^2\over-p^2}\right)
   \right].
\cr
}
\eqn\eight
$$
When $z>0$, the insertion of this function into a Feynman integral
produces new UV divergences besides the conventional UV divergence,
which will turn to be the Borel singularity due to UV renormalons.

With the Borel representation~\seven, the self-energy part
in~Fig.~\figtwo\ is expressed as
$$
\eqalign{
   \Sigma_{ij}^{(2)}(p^2)
   &=-{\delta_{ij}\over N}\int_0^\infty dz\,e^{-z/\lambda}
   \int{d^4k\over i(2\pi)^4}{1\over m^2-(k-p)^2}B(z;k^2)
\cr
   &=-{\delta_{ij}\over N}\int_0^\infty dz\,e^{-z/\lambda}
   F(z)p^2+\cdots.
\cr
}
\eqn\nine
$$
We consider only the wave function renormalization part. Similarly,
the four point vertex function in~Fig.~\figthree\ with the zero
external momentum is given by
$$
   \Gamma_{ijkl}^{(2)}(0)={1\over N^2}(\delta_{ij}\delta_{kl}
   +\delta_{ik}\delta_{jl}+\delta_{il}\delta_{jk})
   \int_0^\infty dz\,e^{-z/\lambda}G(z),
\eqn\ten
$$
where the Borel transform of the vertex function is given by
$$
\eqalign{
   G(z)
   &=i\Delta(0)\int{d^4k\over i(2\pi)^4}{1\over(m^2-k^2)^2}B(z;k^2)
\cr
   &\quad+2\int_0^zdw\,
   \int{d^4k\over i(2\pi)^4}{1\over(m^2-k^2)^2}B(z-w;k^2)B(w;k^2).
\cr
}
\eqn\eleven
$$
{}From \eight, \nine\ and~\eleven, we see that the analytic
continuation of the Borel transforms $F(z)$ and~$G(z)$ has pole
singularities at $z=2n/\beta_1$ for non-negative integer~$n$. The
singularity at $z=0$ corresponds to the usual UV divergence and thus
is removed by the conventional renormalization. The generic UV
renormalon contribution corresponds to~$n\geq1$.

Let us now concentrate on the leading UV renormalon which contributes
to the leading large order behavior and the leading
``non-perturbative'' correction for a weak coupling. It emerges as
the singularity of $F(z)$ and~$G(z)$ at~$z=2/\beta_1$, i.e., the
closest singularity from the origin of the Borel $z$~plane. To study
this singularity, it is sufficient to retain the $O(m^2/p^2)$~terms
in~\eight, because higher order terms give UV convergent integrals
for~$z\sim2/\beta_1$. Then, from elementary calculations, we see that
the analytic continuation of~$F(z)$ behaves near~$z=2/\beta_1$ as
$$
   F(z)
   =-{1\over16e^2\pi^2\beta_1}{m^2\over\mu^2}{1\over z-2/\beta_2}
   +\cdots,
\eqn\twelve
$$
and, similarly, $G(z)$ has the double pole,
$$
\eqalign{
   G(z)
   &={1\over e^2\pi^2\beta_1^3}{m^2\over\mu^2}
   \left({1\over\ln\Lambda^2/m^2}+2\right){1\over(z-2/\beta_1)^2}
\cr
   &\quad+{1\over2e^2\pi^2\beta_1^2}{m^2\over\mu^2}
   \left[\left({1\over\ln\Lambda^2/m^2}+2\right)
         \left(\ln{m^2\over\mu^2}-1\right)+2\right]
   {1\over z-2/\beta_1}
   +\cdots.
\cr
}
\eqn\thirteen
$$
In terms of the large order behavior, the double and the single pole
singularities give rise to $(\beta_1/2)^nn!\lambda^n$
and~$-(\beta_1/2)^n(n-1)!\lambda^n$~behavior respectively, and the
coefficient is given by the residue.

Therefore, by retaining the contribution of the leading UV renormalon
in~\nine, the full propagator of~$\phi_i$ is given by (to this order
of the $1/N$~expansion)
$$
\eqalign{
   iS(p^2)
   &=-\left[\delta_{ij}(p^2-m^2)+\Sigma_{ij}^{(2)}(p^2)\right]^{-1}
\cr
   &=-\delta_{ij}\left[\left(1
   \pm{1\over N}{i\over16e^2\pi\beta_1}{m^2\over\Lambda^2}\right)
   p^2+\cdots\right]^{-1},
\cr
}
\eqn\fourteen
$$
where $+$ ($-$)~sign is for the upper (lower) integration contour
around the pole singularity. It might be possible to specify which is
the physically correct one by tracing back the effect of the
$i\varepsilon$~prescription (Feynman's boundary condition), as was
stressed in~[\OLE]. From~\fourteen, we see that the canonical
normalization of the kinetic term requires the ``wave function
renormalization'' due to the renormalon,
$$
   {\phi_i}_{\rm eff}=\left(1
   \pm{1\over N}{i\over32e^2\pi\beta_1}{m^2\over\Lambda^2}\right)
   \phi_i.
\eqn\fifteen
$$
The four point function is similarly given by only retaining the
lowest $1/N$~expansion and the leading renormalon contribution
in~\ten,
$$
\eqalign{
   &\Gamma_{ijkl}^{(1)}(0)+\Gamma_{ijkl}^{(2)}(0)
   =-{1\over N}(\delta_{ij}\delta_{kl}+\delta_{ik}\delta_{jl}
    +\delta_{il}\delta_{jk})
\cr
   &\times\left[{2\over\beta_1\ln\Lambda^2/m^2}
   \mp{1\over N}{i\over e^2\pi\beta_1^2}{m^2\over\Lambda^2}
   \left(\ln{\Lambda^2\over m^2}+{1\over2}+{1\over2\ln\Lambda^2/m^2}
         \right)\right].
\cr
}
\eqn\sixteen
$$

Now, for a {\it fixed\/} mass parameter~$m$, the contribution of the
UV renormalon in~\sixteen\ can be removed by a finite
renormalization. However, the response of the effective coupling on a
variation of the mass parameter may be of interest in a certain
situation. From~\sixteen\ and the wave function
renormalization~\fifteen, we can read off the effective coupling,
i.e., the four point interaction at the zero momentum, as
$$
   \lambda_{\rm eff}(m^2)=
   {2\over\beta_1\ln\Lambda^2/m^2}
   \mp{1\over N}{i\over e^2\pi\beta_1^2}{m^2\over\Lambda^2}
   \left(\ln{\Lambda^2\over m^2}+{5\over8}+{1\over2\ln\Lambda^2/m^2}
         \right).
\eqn\seventeen
$$
Then the response of the effective coupling on the variation of the
mass parameter might be expressed by a ``$\beta$~function'' (for
$\Lambda\gg m$),
$$
   m{\partial\over\partial m}\lambda_{\rm eff}
   =\beta_1\lambda_{\rm eff}^2
   \mp{1\over N}{4i\over e^2\pi\beta_1^3}
   {1\over\lambda_{\rm eff}}
   e^{-2/(\beta_1\lambda_{\rm eff})}
   +O\left(e^{-4/(\beta_1\lambda_{\rm eff})}\right).
\eqn\eighteen
$$
Therefore we conclude that, to this order of the $1/N$~expansion,
the UV renormalon gives rise to the non-perturbative contribution to
the dimension-four operator and the effective dimensionless coupling.

We have illustrated a non-trivial contribution of the UV renormalon
to a dimension-four operator in a very simple model. However, it is
quite natural to suppose that this phenomenon is universal in
asymptotically non-free theories with a mass parameter or scale. For
example, a similar calculation should be possible in the massive QED,
although the necessary computation will be quite involved.
Presumably, the explicit formula of the two loop vacuum polarization
tensor of massless QED in~[\BEN] will be useful.

After observing the ``non-perturbative'' effect of the UV renormalon,
it seems interesting to examine the possible implication in the
{\it magnetic\/} or {\it dual\/} picture of the Seiberg-Witten theory
on $N=2$~supersymmetric gauge theory~[\SEIB,\REVI]. In the electric
or original picture, the theory is a non-Abelian gauge theory and the
non-perturbative corrections to the effective coupling are provided
by the instantons~[\SEI,\SEIB,\FIN--\AOY]. In the magnetic picture,
on the other hand, the low energy effective theory is described by an
asymptotically non-free Abelian gauge theory with a massive ``dual
electron'' ($=$~magnetic monopole). Therefore the dual theory
fulfills the necessary conditions for the non-trivial contribution of
UV renormalons. Furthermore, to our knowledge, the physical origin of
the non-perturbative corrections in the magnetic picture has not been
fully clarified.

According to the exact solution of~[\SEIB], the effective coupling in
the magnetic picture behaves as\foot{%
It should be noticed that $\Lambda$~in this expression is the
$\Lambda$~parameter in the original $N=2$~supersymmetric su(2) gauge
theory~[\SEIB], and is not necessarily the same as the
$\Lambda$~parameter of the dual (or magnetic) theory, $\Lambda_D$. In
general, $\Lambda$ and~$\Lambda_D$ are proportional to each other.}
$$
   {1\over\alpha_{\rm eff.}(m)}=\Im\tau_D(a_D)
   ={\beta_1\over2}\left(\ln{\Lambda^2\over|m|^2}+c\right)
    +{\beta_1\over2\sqrt{2}}{\Im m\over\Lambda}
    +O\left({m^2\over\Lambda^2}\right),
\eqn\nineteen
$$
near the singular point of the moduli space where the monopole
becomes massless; the mass of monopole~$m$ is related to the VEV of
the dual scalar field~$a_D$ by~$m=\sqrt{2}a_D$. In~\nineteen,
$\alpha_{\rm eff.}=g_D^2/(4\pi)$ is the effective coupling constant
in the {\it dual\/} picture and $\beta_1=1/\pi$ is the one loop
coefficient of the $\beta$~function of~$\alpha_{\rm eff.}$, and
$c=6.23832$~is a constant. The last term of~\nineteen\ is the first
non-perturbative correction to the effective coupling. Accordingly,
the ``$\beta$-function'' is given by,
$$
   |m|{\partial\over\partial|m|}\alpha_{\rm eff.}
   =\beta_1\alpha_{\rm eff.}^2
    -{\beta_1e^{c/2}\over2\sqrt{2}}\alpha_{\rm eff.}^2
    e^{-1/(\beta_1\alpha_{\rm eff.})}\sin\arg m
   +O\left(e^{-2/(\beta_1\alpha_{\rm eff.})}\right).
\eqn\twenty
$$

At first glance, the structure of \nineteen\ and~\twenty\ is quite
resemble to the UV renormalon's effect \seventeen\ and~\eighteen.
However, unfortunately, there are strong arguments that the
non-perturbative corrections in \nineteen\ and~\twenty\ cannot be
regarded as the effect of the UV renormalon: Firstly, if the low
energy effective theory analyzed in~[\SEIB] is the Wilsonian one,
there must exist an UV momentum cutoff~$M$ in the effective theory.
Then UV renormalon cannot emerge because the Landau pole is far
outside the momentum integration region, $\Lambda\gg M$, for a weak
coupling. Even if the effective action in~[\SEIB] is the conventional
one (the generating functional of 1PI Green's functions), it seems
impossible to interpret the non-perturbative corrections in the dual
picture as the UV renormalon's effect. The point is that the
non-perturbative corrections in \nineteen\ and~\twenty, being
depending on the {\it phase\/} of~$m$ and $a_D$, is breaking the
R-symmetry of the dual theory ($N=2$~supersymmetric QED).
Even when the dual scalar field~$a_D$ acquires the VEV and monopole
becomes massive, the mass squared of the monopole is given by~$|m|^2$
which is invariant under the R-rotation. In fact, it can be confirmed
that Feynman diagrams contributing to~$\alpha_{\rm eff.}$ depend only
on a combination~$|m|^2=2|a_D|^2$---the perturbative correction
itself cannot break the global R-symmetry unlike the instanton.
Therefore, the corrections due to UV renormalons, even if there
exist, cannot contribute to the dual {\it holomorphic\/}
prepotential.\foot{%
The one loop perturbative correction is exceptional in this sense,
because the logarithm can be expressed as a sum of holomorphic and
anti-holomorphic functions. Namely, the dual holomorphic
$\tau$-parameter, $\tau_D=i\beta_1\ln\Lambda/m$, reproduces
the one loop correction,
$1/\alpha_{\rm eff.}=\Im\tau_D=\beta_1(\ln\Lambda^2/|m|^2)/2$.}
Clearly, this conclusion is in accord with the non-renormalization
theorem of $N=2$~supersymmetric theory stated in~[\SEI].

This work was supported in part by the Ministry of Education
Grant-in-Aid for Scientific Research, Nos.~08240207, 08640347,
and 08640348.
\refout
\nobreak
\figout
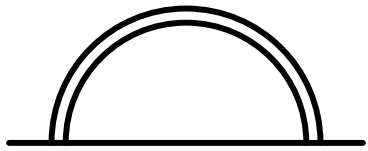
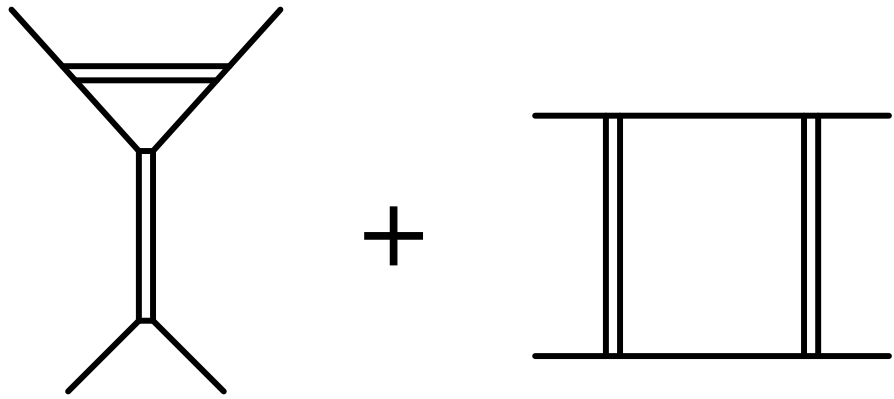
\bye